\documentclass{aastex62}

\submitjournal{PASP}
\received{May 13, 2019}
\revised{May 31, 2019}
\accepted{June 8, 2019}
\submitjournal{PASP}

\shorttitle{Gas Kinematics in NGC 1415}
\shortauthors{Garcia-Barreto et al.}

\begin{document}
\title{H$\alpha$ Nuclear Geyser (Bipolar Outflow) from the Barred Galaxy NGC~1415}
\correspondingauthor{J. Antonio Garcia-Barreto}
\email{j.antonio.garcia.barreto@astro.unam.mx}

\author{Jos\'e Antonio Garcia-Barreto}
\affil{Instituto de Astronomia,
Universidad Nacional Aut\'onoma de M\'exico, \\
Apartado Postal 70-264, Ciudad de M\'exico, M\'exico}

\author{Y. Divakara Mayya}
\affiliation{Instituto Nacional de Astrofisica Optica y Electr\'onica \\
Calle Luis Enrique Erro \# 1, Tonanzintla, Puebla, M\'exico, C.P. 72840}

\author{Jos\'e Guichard}
\affiliation{Instituto Nacional de Astrofisica Optica y Electr\'onica \\
Calle Luis Enrique Erro \# 1, Tonanzintla, Puebla, M\'exico, C.P. 72840}

\begin{abstract}
A long slit spectrum from the barred galaxy NGC 1415 has been obtained with the 2.1m Guillermo Haro telescope in Cananea, M\'exico at position angle $\sim 155\degr$ and shows the kinematics of Na I D lines (in absorption) and H$\alpha \lambda 6562.8$ \AA, [NII] $\lambda 6548$ \AA, [NII] $\lambda 6584$ \AA, [SII] $\lambda 6716$ \AA, and [SII] $\lambda 6731$ \AA~ lines in emission from the central regions and the disk. Our previous H$\alpha$ continuum-free imaging of the central region showed mainly two central bright H$\alpha$ knots straddling the nucleus, and H$\alpha$ emission regions along the south-east and north-west inner spiral arms. Velocities of the Na I D absorption lines are taken as representative of the rotation curve of NGC 1415. Our kinematical data indicates that the central bright H$\alpha$ straddling the nucleus have velocities in excess of the Na I D velocities. We interpret these velocity excesses of the central bright H$\alpha$ knots as due to a geyser (bipolar outflow) with V$_{outflow} \sim 140$ km s$^{-1}$ at a P.A.+165$\degr$. The axis of this outflow, is not along the rotation axis of the disk of NGC 1415 (if it were, it would be at P.A.+238$\degr$). Additionally we have determined $\Omega_{gas}$, radial resonances $\kappa(R)$ and estimated the value of the pattern angular speed of an inner boxy stellar bar in NGC 1415, $\Omega_{bar}$, from the Na I D rotation curve assuming $\mathcal{R} = 1$, $\Omega_{bar} \sim 134$ km s$^{-1}$.
\end{abstract}


\section{Introduction} \label{sec:intro}
To understand the kinematics of the H$\alpha$ bright regions straddling the nucleus of the barred galaxy NGC 1415 and the gas in the inner $30\farcs0$ region, we have performed long slit optical spectral line observation at a PA$\sim 155\degr$ (very close to the PA$\sim 148\degr$ of the disk) mainly including the red portion of the optical spectrum detecting Na I  D (unresolved $\lambda$ 5895.92, 5889.95 \AA) lines, in absorption, H$\alpha$ ($\lambda$ 6562.8 \AA), [N II] ($\lambda$ 6548, 6584 \AA), and [S II] ($\lambda$ 6717, 6731 \AA) in emission.

In our original survey of H$\alpha$ emission from strong barred galaxies, within the Revised Shapley Ames Catalog, with IRAS f(IRAS)$_{60} \geq 5$ Jy, and colors characteristic of star-forming galaxies, we included the disk barred galaxy NGC 1415 \citep{gar96}. Our optical red continuum, filter {\it I} $\lambda 8040$ \AA, observation of the inner $115\farcs0$ of the barred galaxy NGC 1415 shows elongated and boxy-shaped isophotes in the region around the nucleus \citep{gar96,gar00}. Figure 1 is a reproduction of our filter {\it I} $\lambda 8040$ \AA, image of the inner 115$''$ regions of NGC 1415 (in contours). The continuum optical red (filter I) had not been flux calibrated, so the isophotes are in arbitrary units proportional to the equivalent of noise \citep{gar96,gar00}. Figure 2 shows the H$\alpha$ continuum-free images of the inner region of NGC 1415. Left plot shows the emission in contours in arbitrary units proportional to the equivalent of noise \citep{gar96,gar00} Right plot shows the H$\alpha$ continuum-free emission (in grey scale) superposed on the optical red continuum (filter I, in contours). The letter $\mathcal{A}$ indicates the SE bright H$\alpha$ knot, and letter $\mathcal{B}$ indicates the NW bright H$\alpha$ knot.

NGC 1415 (ESO 482-G033) is classified as an SBa in the Revised Shapley Ames catalog \citep{san87}, as RSXS0 in RC3 \citep{dev93}, as (R)SAB0/a(s) in homogenized NED, and as (RL)SAB$_a$(r'l,nr)0$^+$ in the paper Near-IR atlas of S0-Sa galaxies \citep{lau11},  see Table 1. NGC 1415 has galactic coordinates $b \sim 215\degr.7$ and $l \sim -51\degr.4$, namely, it is in the third quadrant of our galaxy and far below from the plane of our galaxy. The values of galactic extinction are $A(R)_V \sim 0.052$ in R band, and $A(I)_V \sim 0.036$ in I band \citep{sch11}. NGC 1415 is a member of a nearby poor cluster, the Eridanus Group \citep{wil89,oma05}, which consists of 54 galaxies within the approximate limits of 3$^h~17^m$ and 4$^h~02^m$ in right ascension (J2000) and -25$\degr~49'$ and -14$\degr~52'$ in declination (J2000) \citep{wil89,oma05}. Twenty five out of 32 S galaxies in the Eridanus group are barred, Hubble type SB, that is 78\%. This is a high percentage of barred galaxies and it might indicate the physical conditions for galaxy formation and evolution on that part of the nearby universe. For the Eridanus group with $\sigma\simeq 265$ km s$^{-1}$ \citep{wil89}, the predicted X~ray luminosity would be L$_X(Eridanus)\sim 4\times10^{43}$erg s$^{-1}$ and the predicted X~ray temperature would be T$_X(Eridanus)\sim 1.12$ keV or T$_X(Eridanus)\sim 1.3\times10^7$ K \citep{edg91}. Radio continuum radiation has been detected from VLA maps from NGC 1415 at 20cm \citep{con90}.
	
NGC 1415 has 2 galaxy companions within 10 diameters (less than 200 kpc away) and 5 galaxy companions within 10 and 20 diameters (between 200 kpc and 350 kpc away, \citep{gar03}.

\begin{figure}[ht!]
\includegraphics[width=15cm,height=19cm]{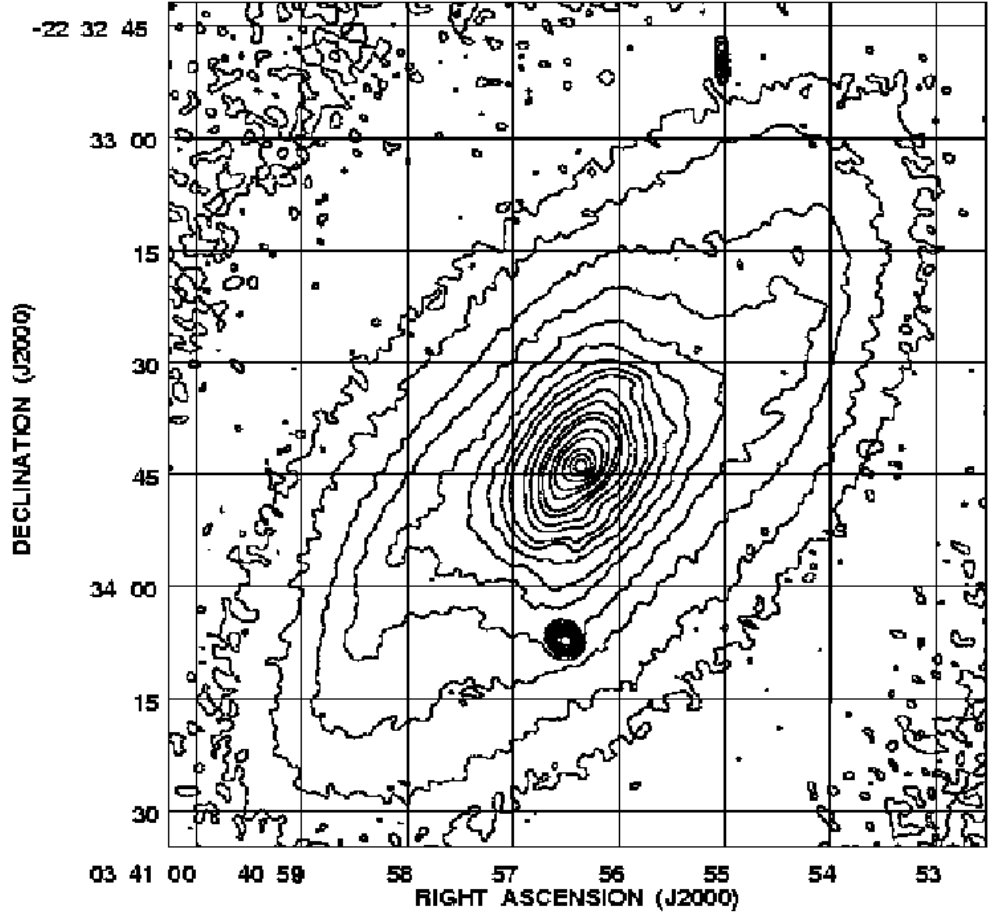}
\figcaption{Our red optical continuum image of the inner $\sim 115\farcs$ region of NGC 1415 in the broadband filter {\it I} (${\lambda}_{central} \simeq~8040$ \AA) \citep{gar96,gar00}. Contours are in arbitrary units proportional to the equivalent noise units (first contour is at 3$\sigma$). Notice the elongated isophotes of the stellar inner inner bar (just outside the nucleus) and the contours just outside the inner inner bar forming an inner boxy shape.  \label{fig. 1}}
\end{figure}
	
Long exposure photographs of NGC 1415, like those in the Hubble Atlas \citep{san62}, indicate a rectangular bar at a PA $\sim 130\degr$ with two main inner spiral arms originating from the SE and NW.  Here we adopt a distance of to NGC 1415 as D$_{N1415} = 17.7$ Mpc (H$_{\circ}=75$ km s$^{-1}$~Mpc$^{-1}$, \citep{tul88}), the linear scale is $1\farcs0 \sim 85.81$ pc. Numerical N-body simulations of disk galaxies have shown that they may develop a high eccentricity bar with an elliptical shape (Martin 1995; Martin \& Friedli 1997). Sparke \& Sellwood (1987), Combes et al. (1990) and Athanassoula et al. (1990) have shown, however, that bars in some disk galaxies are more rectangular than elliptical.

\begin{figure}[ht!]
\includegraphics[width=6cm,height=8cm]{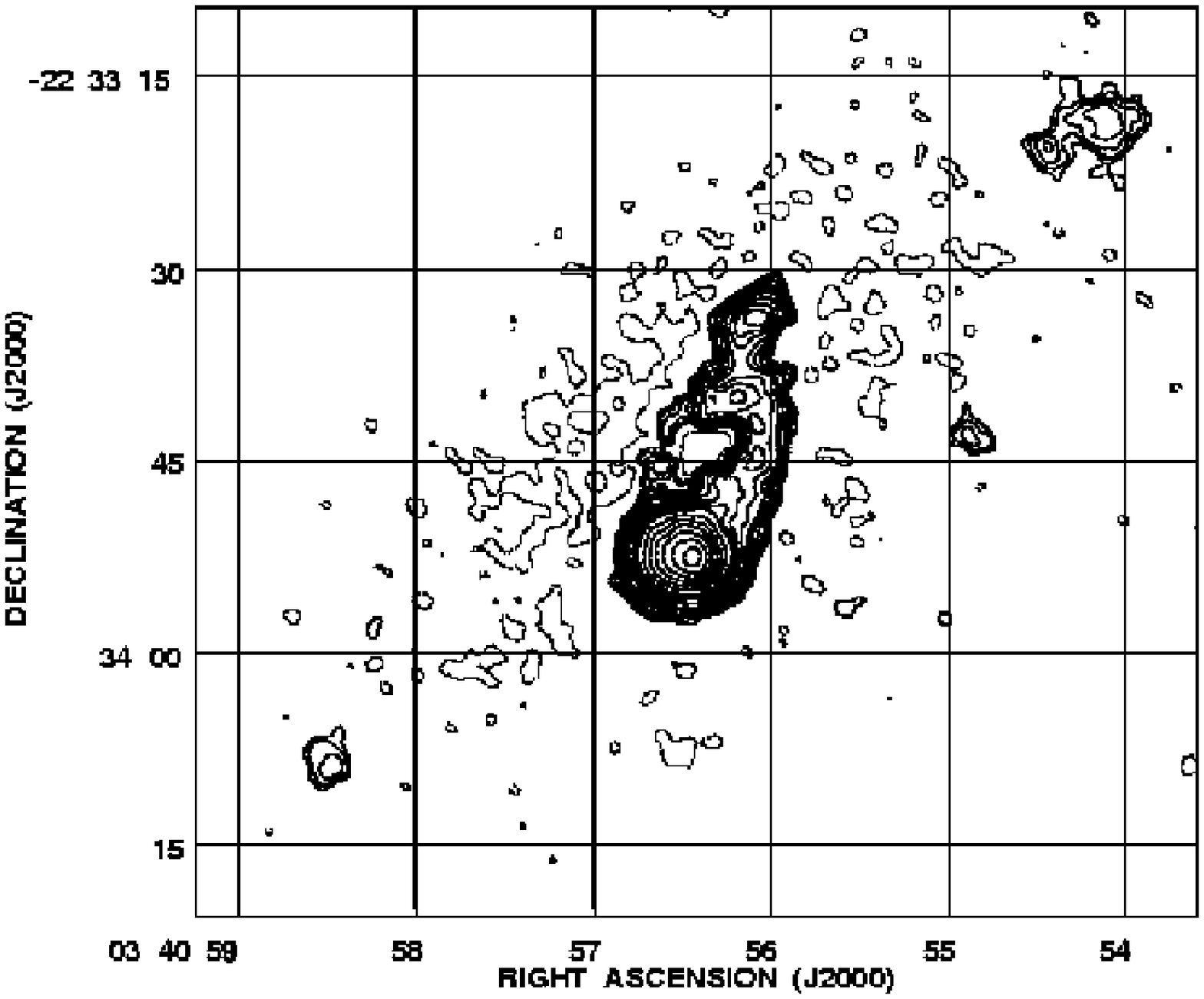}
\includegraphics[width=8cm,height=8cm]{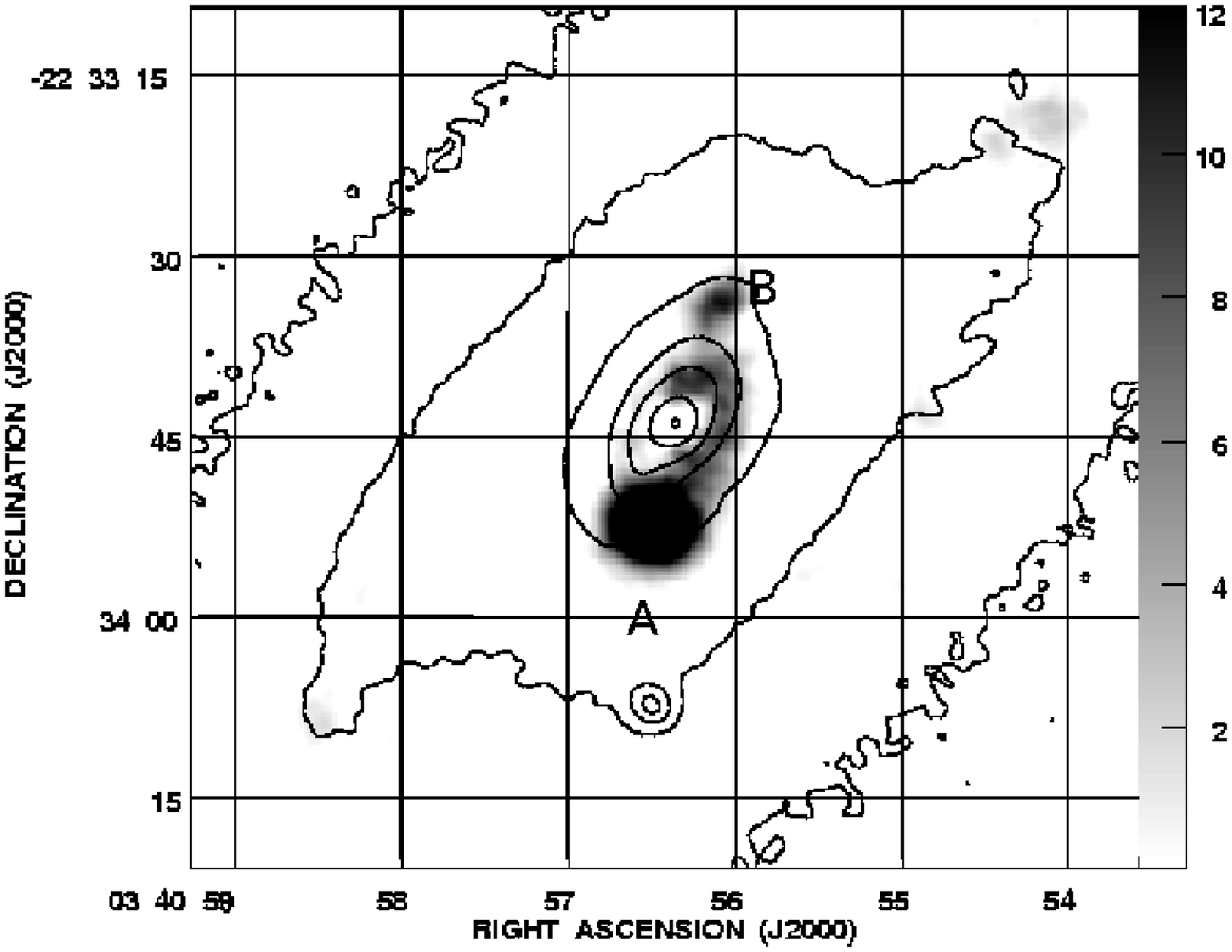}
\figcaption{Left plot shows our H$\alpha$ continuum-free image (in contours, in arbitrary units proportional to the equivalent noise units with first contour at 3$\sigma$) \citep{gar96,gar00}. Right plot shows  the same as in left plot, now in grey scale superposed on our red optical continuum image (I filter, in contours) with the letter $\mathcal{A}$ indicating the south-east (SE) bright H$\alpha$ knot, and with the letter $\mathcal{B}$ indicating the north-west (NW) less bright H$\alpha$ knot \citep{gar96,gar00}. \label{fig. 2}}
\end{figure}
	
Important for the present study of the barred galaxy NGC 1415 are: {\bf a)} the position of the optical nucleus $\alpha(J2000.0)$ and $\delta(J2000.0)$ of the galaxy. {\bf b)} the relative spatial positions and distances of the structures observed in the H$\alpha$ gas continuum-free optical brightness distribution (from our optical observations \citep{gar96,gar00}).  {\bf c)} The systemic velocity of NGC 1415.  {\bf d)} The inclination of the disk of NGC 1415 with respect to the plane of the sky. See Table 1 for adopted parameters.
	
In this paper we anchored the position of the nucleus in our red optical continuum (filter I), and in our H$\alpha$, continuum-free image, with the optical position of the nucleus (from palomar plates \citep{gal75}, see Table 1).
	
The barred galaxy NGC 1415 appears to be one of several {\it normal} disk galaxies with a geyser (or bipolar outflow) from the nucleus, see $\S$6 below.	

In this paper, we present the observed kinematics of the gas distribution in the central region of the galaxy using primarily the Na I D absorption lines, and H$\alpha$ emission lines from a long slit spectrum along P.A.$\sim 155\degr$. In $\S$ 2 we describe the long slit optical spectral observation. In $\S$ 3 we describe the surface brightness distribution of the optical red continuum , filter {\it I $\lambda 8040$ \AA} and the H$\alpha$ continuum-free. In $\S$ 4 we describe the kinematics of the Na I D absorption lines, and the H$\alpha$ emission lines, and we describe our estimate for the rotation curve from the unresolved Na I D absorption lines at a P.A.$\sim 155\degr$ (E of N) using the redshifted and blueshifted velocities from the systemic heliocentric velocity of NGC 1415. In $\S$ 5 we describe the high values of the H$\alpha$ velocities from $\mathcal{A}$, and $\mathcal{B}$ knots. In $\S$ 6 we compute, from the rotation curve, $\Omega_{gas}$, the radial epicycle frequencies $\kappa(r)$, estimate the values of $\Omega_{gas} - \kappa/2$, $\Omega_{gas} + \kappa/2$, and estimate the angular velocity $\Omega_{bar}$ for the boxy stellar bar in NGC 1415. In $\S$ 7  we summarize our findings.

\clearpage
\newpage
\startlongtable
\begin{deluxetable}{lcc}
\tablecaption{General Properties of the barred disk galaxy NGC 1415 \label{tab:table}}
\tablehead{
\colhead{Characteristic} & \colhead{Value} & \colhead{Reference} \\
}
\colnumbers
\startdata
Hubble Type (RSA)  & SBa                                     & 1 \\
Distance           & 17.7 Mpc                                & 2 \\
Spatial scale      & $1\farcs0 \sim 85.81$ pc                                & 2 \\
IRAS $12\mu$       & 0.26 (0.53)$^a$ Jy                      & 3, 4, 5 \\
IRAS $25\mu$       & 0.55 (0.53)$^a$ Jy                      & 3, 4, 5 \\
IRAS $60\mu$       & 5.28 (6.72)$^a$ Jy                      & 3, 4, 5 \\
IRAS $100\mu$      & 12.32 (12.64)$^a$ Jy                    & 3, 4, 5 \\
T$(IRAS)_{dust}$   & 33$\degr$ K                          & 4 \\
log L$_B/L_{\odot}$ &  9.73                                  & 2 \\
Heliocentric Systemic Velocity  &  V$(hel)_{sys} \sim 1564 \pm 9$ km s${-1}$ & 6,7 \\
Photometric major axis P.A. & 148$\degr E of N $          & 8\\
Stellar boxy bar P.A.&  130$\degr\pm 5$                   & 4, 9 \\
Stellar boxy bar semi major diameter &  17$\farcs5 = 1.5$ kpc   & 9 \\
M(HI)              & $5.75 - 7.1\times10^8$ M$_\odot$        & 10 \\
Total radio continuum (20 cm) $\theta_{fwhm}\sim 8''$, & S$_T\sim 18$ mJy & 11 \\
Peak SE knot radio continuum (20 cm) $\theta_{fwhm}\sim 8''$, S$_p$, & 6.1 mJy & 11 \\
Right Ascension (J2000) & $3^h ~~ 40^m ~~ 56^s.921$           & 12 \\
Declination (J2000) & $-22\degr ~~ 33' ~~ 49\farcs.507$           & 12 \\
Inclination of the disk $i$ & 65$\degr$                      & 13 \\
\hline
\enddata
\tablenotetext{a}{Total IRAS fluxes using ADDSCAN (Rush et al. [1993]}
\tablecomments{1) RSA Sandage \& Tammann, (1987), 2) Tully (1988), 3) IRAS Point Source Catalog, 4) Garcia-Barreto et al. (1996), 5) Rush et al. (1993),  6) Lauberts (1982), 7) Da Costa et al. (1998), 8) de Vaucouleurs et al., RC3 (1993) 9) Garcia-Barreto \& Moreno (2000), 10) Huchtmeier (1982),  11) Condon et al. (1990), 12) Gallou\"{e}t et al. (1975)}
13) Tully (1988) \end{deluxetable}

\section{Optical Spectral Observations} \label{sec:observations}

An optical line emission spectrum was obtained at the 2.1m optical telescope of the Observatorio Astrofisico Guillermo Haro (OAGH) in Cananea, Sonora, M\'exico, operated by the Instituto Nacional de Astrofisica Optica y Electr\'onica (INAOE), during the nights of December 29 and 30, 2000. The observatory location is latitude +31$\degr$ 03$'$ 10$''$ and longitude 110$\degr$ 23$'$ 05$''$ west at an altitude of 2480m above the mean sea level. A grating with 300 groves mm$^{-1}$ was used which resulted in a spectral sampling of about 7.6 \AA~  and spectral coverage of $\approx 5000 \rightarrow 6800$ \AA. The spectral and spatial samplings were 1.66 \AA~  pixel$^{-1}$ and $0\farcs463$ pixel$^{-1}$, respectively. Slit width was $\sim 1\farcs8$ with lenght of about 80$''$. The slit P.A. was $\sim 155\degr$ close to the P.A.$\sim 148\degr$ of the major axis of the optical disk. The air mass was 1.75. Three spectra of 30 minutes each were taken. The instrumental response was calibrated by the observation of the standard star Feige 25. Wavelength calibration was established via observations of He/Ar lamp \citep{may04}.

Each frame was bias-corrected and divided by a normalized flat field, using various tasks in the IRAF optical image analysis package. Three wavelength-calibrated frames at each P.A. were averaged ({\it task imcombine in IRAF}), in the process of removing cosmic-ray events. Sky spectrum was extracted from the slit from the object-free regions and subtracted from the NGC 1415's spectrum. The uncertainity of the peak of the emission line determination was about 10\%, or about $\pm 0.38$ \AA~ (this would correspond to about $\pm 17$ km s$^{-1}$ at H$\alpha$, and to about $\pm 19$ km s$^{-1}$ at Na I D$_1$ line $\lambda 5895.92$ \AA).
	
We were able to detect the neutral gas through sodium (Na I D lines, see appendix) in absorption throughout the disk of NGC 1415, in the slit. Our long slit spectrum observation was unable to resolve each Na I D$_1$ and D$_2$ lines ($\lambda_{D2} \sim 5889.951$, and $\lambda_{D1} \sim 5895.924$) in velocity, since our instrumental spectral sampling was about 7.6 \AA. Thus we fitted gaussian profiles for the combined symmetrical absorption Na I D lines and determine their peak velocities in space and velocity from the spectrum. In the case of our kinematical study of the gas in NGC 1415, the Na I D lines are also redshifted (similar to the redshifted of emission lines \citep{gra79}) and thus one confidently may assume that the absorbing gas shares the orbital rotational motion of the gas around the center of the disk.

\section{Surface Brightness Spatial Distribution in NGC 1415} \label{sec:surface}
\subsection{Red Optical Continuum, filter {\it I} ($\lambda 8040$ \AA)}

Figure 1 shows a reproduction of our optical red continuum image (innermost $\sim 115''$ region) of NGC 1415, {\it filter I $\lambda_c\sim 8040$ \AA, $\Delta \lambda\sim 1660$ \AA} from our survey work of bright and nearby RSA barred disk galaxies (with the criteria f$(IRAS)_{60\mu} \geq 5$ Jy) where details about the observations and the data reduction can be found \citep{gar96}. The image was not flux calibrated (the contours are in arbitrary units of relative intensity (r.i.) of the equivalent noise, starting with first contour at 3$\sigma$) \citep{gar96,gar00}.

From our previous analysis, four basic mass structures were semi analytically modeled to fit the relative intensities of the isophotal observations of the optical red continumm \citep{gar00}, namely, (1) a compact bulge of radius $\sim 300$ pc, (2) an elongated elliptical inner inner stellar bar of radius ${\sim}11.6''$ ($\sim 1$ kpc), at a P.A. $\sim 150\degr$, (3) a inner boxy contours outside the inner inner elliptical bar delineating an inner boxy bar, with radius $\sim 17\farcs.5$ (1.5 kpc), and (4) a disk with semimajor axis $\sim 9$ kpc \citep{gar00}. Notice that the boxy bar is not seen edge-on, but at an inclination of $i \sim 65\degr$. The outer spiral arms in our optical red image start from the boxy-shaped isophotal contours radius $\sim 18\farcs.6$ (1.6 kpc) in the NW and in the SE directions suggesting (if they trail) that the disk galaxy is rotating clockwise.

\subsection{Structure decomposition of optical (BVRI), $K_s$, and UV continuum images from the literature}

Two dimensional structural surface brightness BVRI decomposition of 605 galaxies with $B_T \leq 12.9$ mag and $\delta \leq 0\degr$ has been done by the Carnegie-Irvine group \citep{ho11,li11}. NGC 1415 is among the galaxies that they analyzed. They report among other results, B, V, R, and I magnitudes, radial profiles, color index maps, isophotal and photometric parameters, radius enclosing 20\%, 50\%, and 80\% of the light in the B band, ellipticity, $e$, and P.A. of the photometric major axis in the I band, the inclination angle \citep{ho11}. Additionallythey performed Fourier decomposition of the isophotes to quantify non-axisymmetric deviations in the light distribution (namely, bars) \citep{li11}.

They used the {\it task ellipse} in IRAF to determine the values of the $e$ and P.A. of a stellar bar with the following arbitrary criteria to decide if a galaxy has or has not a stellar bar: if none of the points in the $e$ profiles exceeds 0.2 or if $\Delta e \leq 0.1$ throughout the entire $e$ profile they classified a galaxy as non barred \citep{li11}. On the other hand if $e \geq 0.2$ and P.A. is constant then they decide that a galaxy is barred with a length of the bar at a radius when $e$ and P.A. begin to show large deviations ($\Delta P.A. \geq 10\degr$) \citep{li11}. Unfortunately for the case of the galaxy NGC 1415 they report it as non barred \citep{li11}. However, we belive that their result is in error, since the inner stellar bar in NGC 1415 has boxy isophotal shape\footnote{A rectangular, or boxy isophotal bar cannot ever be described by an ellipse, that is, how can a simple ellipse account for the corners of a boxy (rectangular) shape bar? Is there any meaning of $e$ for rectangular isophotes? }.

Two dimensional structural surface brightness of $K_s$-band images of 206 galaxies has been done for S0-Sa galaxies (NGC 1415 is included) in order to report a detailed morphological classification  \citep{lau11}. The decomposition were made by fitting S\'ersic functions for the bulge, an exponential function for the disk and a Ferrers function for the stellar bar \citep{lau11}. For measuring bar lengths, they used two main methods: (1) visual estimation by marking the outskirts of the bar and drawing an ellipse to that distance, (2) radial profiles of the $e$ were used where the bar length was taken to be the radial distance where maximum $e$ in the bar region appeared \citep{lau11}. For NGC 1415 they report a stellar bar of type AB$_a$, P.A.$_{bar}\sim 133\degr$, a$_{bar}^{ell} \sim 27\farcs4$ and $a_{bar}^{visual} \sim 35\farcs0$, $b_{bar}/a_{bar} \sim 0.43$ \citep{lau11}. This NIR $K_s$ bar would include the inner optical stellar bar and the inner optical spiral arms (see Fig. 1)

Two dimensional structural surface brightness of NGC 1415 Spitzer 3.6$\mu$ image has been done \citep{sal15}, modeling a bulge with a radius of $\sim 427$ pc, an exponential disk (with an exponential scale length of $\sim 9.4$ kpc and a P.A.$_{disk}\sim 152$ E of N), and a modified {\it ferrer2} profile for a {\bf only one component} bar, $a_{bar} \sim 1.35$ kpc ($\sim 15\farcs7$ at a P.A.$_{bar}\sim +134$ E of N) \citep{sal15}.

Structure decomposition from UV ($\lambda_{W2} \sim 2030~\AA$ $\lambda_{M2} \sim 2231~\AA$, $\lambda_{W1} \sim 2030~\AA$), and optical U, $\lambda_{U} \sim 3501~\AA$, B, $\lambda_{B} \sim 4329~\AA$, and V, $\lambda_{V} \sim 5402~\AA$, continuum images from {\it Swift}-UVOT has been done for 11 galaxies including NGC 1415 \citep{ram17}. In particular for NGC 1415 they report a nuclear ring ($nr$) with a$_{nr}\sim 10\farcs8$, b$_{nr} \sim 5\farcs2$, and P.A.$_{nr}\sim 166\degr \pm 2\degr$. In their reporting of a nuclear ring from the continuum images, they included the $\mathcal{A}$ and $\mathcal{B}$ knots. They did that without kinematical information. With kinematical information, as discussed in $\S$ 5, $\mathcal{A}$ is moving away from the center (blueshifted), and $\mathcal{B}$ is moving away from the center also (redshifted), and thus we do not think that $\mathcal{A}$ nor $\mathcal{B}$ knots are members of the nuclear ring.

\subsection{Spatial Distribution of H${\alpha}$ Emission Regions}

In NGC 1415 the inner H$\alpha$ emission originates from regions within the inner $\pm 20''$. The H$\alpha$+N[II] continuum-free image was obtained with a set of two narrowband filters with $\lambda \sim 6459$ \AA~with $\Delta \lambda \sim 101$ \AA~ and $\lambda \sim 6607$ \AA~  with $\Delta \lambda \sim 89$ \AA~ \citep{gar96,gar00}. $\mathcal{A}$ indicates (see right plot in Fig. 2) the SE H$\alpha$ (bright by $\sim 3.6$ times, in relative units, than the weaker NW component) component, and $\mathcal{B}$ indicates the NW H$\alpha$ (weak) component.

There are four characteristics, of the H$\alpha$ emission from the central region of NGC 1415, worth emphasizing : {\bf 1)} there are two bright regions straddling the nucleus $\mathcal{A}$ and $\mathcal{B}$ with weak emission connecting to what might be an apparent inner nuclear ring, the P.A. of an imaginary line joining $\mathcal{A}$ and the nucleus and the P.A. of an imaginary line joining the nucleus with $\mathcal{B}$ are  the same $\sim 161\degr \pm 4\degr$ E of N, {\bf 2)} there is H$\alpha$ emission west of the nucleus, which might apparently be part of a circumnuclear structure (or nuclear ring), {\bf 3)} there is no H${\alpha}$ emission from the compact nucleus (assuming the subtraction of the red continuum was done correctly \citep{gar96}), {\bf 4)} there is emission from several regions in the disk.

The P.A. of $\mathcal{A}$ and $\mathcal{B}$ regions is off by $\sim 13\degr \rightarrow 17\degr$ from the P.A. of the major axis of the galaxy P.A.$_{A-B} \sim 161\degr \pm 4\degr$, versus P.A.$_{optical-disk} \sim 148 \pm 4$. With kinematic information (see $\S4$ and $\S5$) and the spatial orientation of the inner spiral arms, the disk is rotating clockwise, assuming that the inner spiral arms are trailing. The peak of the weak H$\alpha$ emission just to the W of the nucleus at PA$\sim 90\degr$ lies at a radius of $4''$ or about 345 pc. The peak of $\mathcal{A}$ H$\alpha$ knot is relatively brighter than the $\mathcal{B}$ knot. The peak of $\mathcal{B}$ lies at a distance of ${\sim}10\farcs3$ or $\sim 884$ pc. $\mathcal{A}$ lies at a distance of $\sim 8\farcs8$ or $\sim 755$ pc.

It is now well accepted that gas in a non-axisymmetric bar potential loses angular momentum when gas inside the corotation radius (CR) is transfered inwards, and could accumulate near an Inner Lindblad Resonance (ILR), while gas outside CR is transfered outwards and could accumulate near an Outer Lindblad Resonance (OLR)\footnote{x$_1$ orbits in a non-axisymmetrical gravitational potential (stellar bar) is one family of orbits that are elongated and sustain the stellar bar along the major axis\citep{con80a,con80b}. x$_2$ orbits is another family of orbits closer to the nucleus but are elongated perpendicular to the x$_1$ orbits along the minor axis of the stellar bar \citep{lyn72,con80a,con80b}.} \citep{lyn72,sch84,com85,bin87}. Density enhancements (with star formation) near an IILR might form a circumnuclear structure (CNS) \citep{sch84,com85}. Other RSA bright and near barred galaxies with clear CNS at r$\sim 6 \rightarrow 9 \farcs0$ are, for example, NGC 1326 \citep{gar91a}, and NGC 4314 \citep{gar91b}.

\begin{figure}[bth]
\includegraphics[width=8cm,height=8cm]{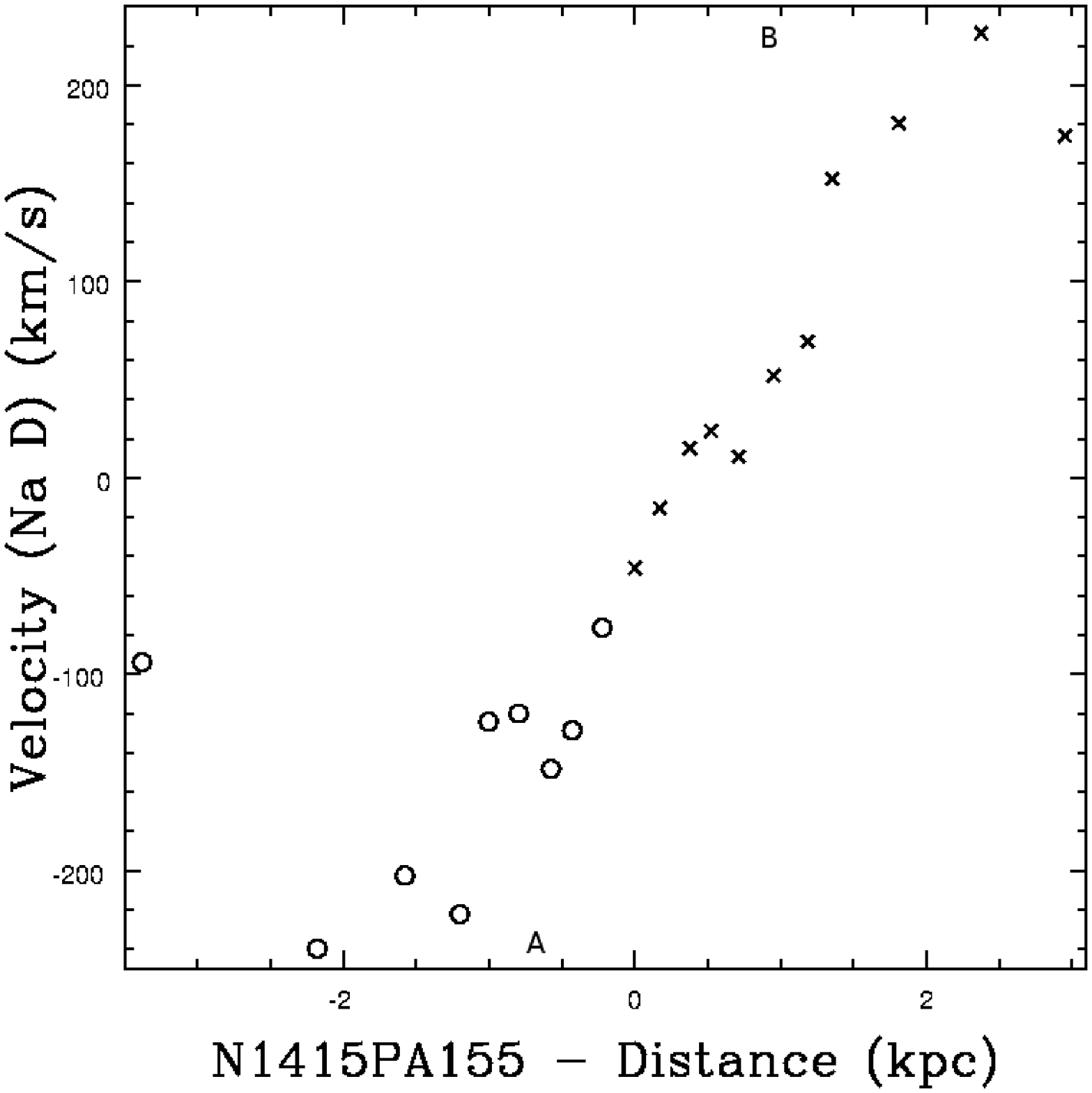}
\includegraphics[width=8cm,height=8cm]{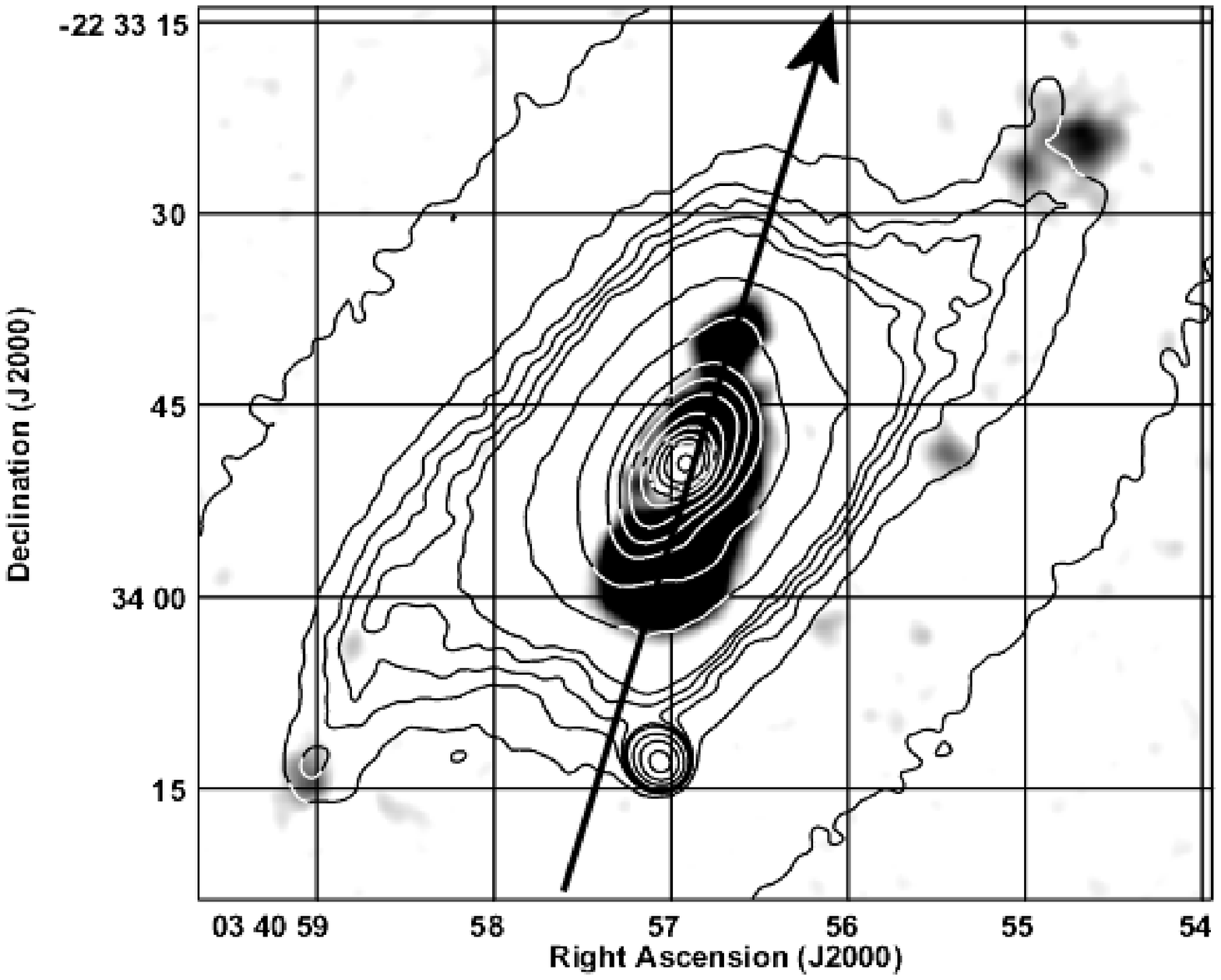}
\figcaption[figure=gdgnew12.eps]{Left plot shows the heliocentric velocities of neutral sodium, Na I D lines, relative to V$_{sys}=1564$, $i=65\degr$, and $\Delta \theta = 7\degr$ along long slit at P.A. +155$\degr$, negative offsets are from SE (open circles), positive offsets are from NW (crosses). Letter B on north axis, and A on bottom axis denote approxmimately the space locations of $\mathcal{B}$ and $\mathcal{A}$ knots. Right plot shows of a line over the optical red continuum (in contours) and H$\alpha$ image (in grey scale) at an approximate position of the long slit at P.A.$=155\degr$ (the arrow indicates the increase in distance).  \label{fig. 3} }
\end{figure}

Without kinematic information two plausibles interpretations might have been considered for the existence of $\mathcal{A}$ and $\mathcal{B}$ regions. First, they might have been regions of density enhancement and recent star formation due to shocks when gas meets the stellar bars in NGC 1415 \citep{ath92} and try to follow x$_1$ orbits along the major axis of the bars. However, as seen in the plane of the sky, the spatial locations of $\mathcal{A}$ and $\mathcal{B}$ are in the expected spatial positions if the stellar bars were rotating counter-clockwise \citep{ath92}. However, as mentioned earlier, from the spatial location, curvature of spiral arms (trailing), and kinematics, the disk rotates clockwise and thus we assume that the bars do also rotate likewise, with the NE side being closer to the observer. Thus this first interpretation is not valid with kinematic information.

Second, $\mathcal{A}$ and $\mathcal{B}$ might be considered knots of a nuclear ring with a$_{nuclear-ring} \sim 10\farcs8$, b$_{nuclear-ring} \sim 5\farcs2$ at P.A.$\sim 166 \pm2 \degr$ \citep{ram17}. This is the interpretation of the continuum UV and optical U, B and V imaging observations but without any kinematical analysis\citep{ram17}.

However with information of the kinematics of the gas (see $\S4$, and $\S5$) $\mathcal{A}$ and $\mathcal{B}$ represent regions that are not in the rotating plane of the galaxy, but instead they form a different system, namely, gas moving away from the nucleus. The imaginary line joining $\mathcal{A}$ and $\mathcal{B}$ is at P.A.$\sim 161\degr \pm 4\degr$, making an angle $\Delta \theta \sim 13\degr \rightarrow 17\degr$ from the P.A.$\sim 148$ of the major axis of the galaxy, and making an angle $\Delta \theta \sim 73\degr$ from the P.A. of the rotation axis of the galaxy (P.A.$_{rotation-axis} \sim 238\degr$). The nuclear ring would have r$_{nuclear-ring} \sim 4\farcs0 \rightarrow 5\farcs0$ \citep{ram17} (see previous paragraph).

\section{Kinematics of Na I D absorption, and H$\alpha$ emission lines in NGC 1415}
\subsection{Velocities of gas in the gravitational field of a disk galaxy}

The velocity of gas under the gravitational potential of a disk galaxy could be expressed in cylindrical coordinates as
	
$$ \vec{V}(R, \theta, z) = V_R \hat{r} + V_{\theta} \hat{\theta} + V_z \hat{z}.$$

For an outside observer, the heliocentric velocity from different regions in the disk of a galaxy  with a heliocentric velocity, V$_{sys}$, inclined an angle $i$ with respect to the plane of the sky, and from an azimuthal angle $\theta$ from the major axis is given by \citep{mih81}
	
$$ V(obs)_{hel} = V(hel)_{sys} + V_R(R, \theta) sin {\it i} sin \theta + V_{\theta}(R, \theta) sin {\it i} cos \theta + V_{z}(R, \theta) cos {\it i}.$$

In the first approximation that gas is on circular orbits, with V$_R(R, \theta)$ = 0, and V$_z(R, \theta)$ = 0, and the above expression becomes

$$ V(obs)_{hel} = V(hel)_{sys} + V_{\theta}(R, \theta) sin ({\it i}) cos (\theta). $$

For a long slit spectrum observation along the major axis of a disk galaxy $\theta = 0\degr$, the velocities are:

$$ V_{\theta}(R, \theta=0)  = \frac {V(obs)_{hel} - V(hel)_{sys}}{sin ({\it i})}, $$

and for a long slit spectral observation at different position angle, the velocities are:

$$ V_{\theta}(R, \theta)  = \frac {V(obs)_{hel} - V(hel)_{sys}}{sin ({\it i}) cos (\Delta \theta)}, $$

where $\Delta \theta$ is the angle difference between the P.A. of the slit and the P.A. of the major axis of the galaxy.

In our long slit spectroscopy study of NGC 1415, we consider V(hel)$_{sys} = 1564$ km s$^{-1}$, {\it i} = 65$\degr$,  P.A.$_{phot-axis} = 148\degr$, and $\Delta \theta_{slit155} = 7\degr$ (the difference in P.A. of the slit and the P.A. of the photometric major axis).

\subsection{Velocity Curve from a slit spectrum at P.A.$\sim 155\degr$}

The NW distances from the center show redshifted velocities, while the SE distances from the center show blueshifted velocities. The north SE-NE edge of NGC 1415 is closer to the observer. Assuming the inner spiral arms are trailing, then the direction of galaxy inner rotation is clockwise. The inner stellar boxy bar in NGC 1415 is at P.A.$_{bar}\sim130\degr$. A line joining $\mathcal{A}$ and $\mathcal{B}$ knots, on the plane of the sky, straddling the nuclear region has P.A.$_{AB}\sim165\degr$.

Neutral sodium, Na I, {\it D lines} correspond to its doublet-resonance transition (see Appendix) are seen in absorption in the interstellar medium \citep{spi78,dys80}. Observations of Na I D lines, in our galaxy, indicate that the diffuse neutral gas is confined to a disk thickness approximately 250 pc \citep{spi78,dys80}. Since, in NGC 1415, Na I D lines (optically thin, cold, and most probably on the plane of the galaxy) are detected in absorption, it is a reasonable assumption that they represent well the kinematics of gas in rotation orbiting around the center of NGC 1415, as was first detected in Cen A \citep{gra79}.

Figure 3 left plot shows the neutral sodium, Na I D line heliocentric velocities respect to V$_{SYS}$ versus distance at a PA$155\degr$ EofN. Negative relative distances, left on figure, correspond to SE of NGC 1415, while positive relative distance correspond to NW of NGC 1415. The velocities shown take into account $\Delta \theta = 7\degr$, they are smoothly rising with abrupt changes. Right plot shows the line at P.A.$\sim +155\degr$ over the red optical continuum (in contours) and H$\alpha$ image (in grey scale) indicating approximately the spatial position of the long slit. Letter B on north axis, and A on bottom axis denote approximately the space locations of $\mathcal{B}$ and $\mathcal{A}$ knots.  Notice that the long slit mainly shows velocities associated with $\mathcal{A}$ knot, the north of a plausible nuclear ring and $\mathcal{B}$ knot. This spectral slit P.A. is very close to the photometric PA$\sim 148\degr$ of the major axis of the disk.



\begin{figure}[bth]
\includegraphics[width=8cm,height=8cm]{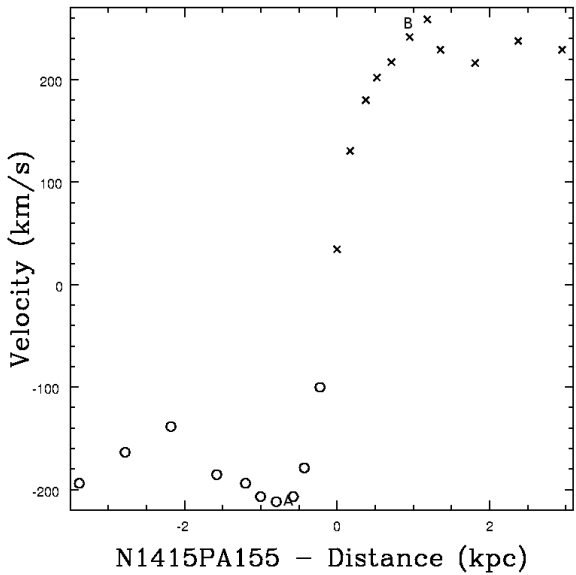}
\figcaption[figure=gdgnew15.ps]{Velocity versus distance for the recombined H$\alpha$ emission at P.A.$\sim 155\degr$ (EofN). Letter B on north axis, and A on bottom axis denote approximately the space locations of $\mathcal{B}$ and $\mathcal{A}$ knots. Notice that the velocities shown  also smoothly rise but with a higher slope and reach V$\sim 200$ km s$^{-1}$ at d$\sim 0.5$ kpc) and V$\sim 260$ km s$^{-1}$ at d$\sim 1.2$ kpc)  \label{fig. 4} }
\end{figure}

Figure 4 shows H$\alpha$ heliocentric velocities versus distance at a P.A. $\sim +155\degr$ (EofN). Negative relative distances, left on plot, correspond to SE of NGC 1415, while positive relative distances correspond to NW of NGC 1415. The SE (negative distances) velocities shown are smoothly rising  and reach V$\sim 100$ km s$^{-1}$ at d$\sim 0.22$ kpc, then from d$\sim 0.43$ kpc the velocities increase with distance and reach V$\sim 211$ km s$^{-1}$ at d$\sim 0.8$ kpc), then slowly decrease to a value (V$\sim 180$ km s$^{-1}$ at d$\sim 1.65$ kpc, continue decreasing to a value (V$\sim 138$ km s$^{-1}$ at d$\sim 2.2$ kpc,  V$\sim 164$ km s$^{-1}$ at d$\sim 2.78$ kpc, and finally V$\sim 194$ km s$^{-1}$ at d$\sim 3.4$ kpc. Letter B on north axis, and A on bottom axis denote approximately the space locations of $\mathcal{B}$ and $\mathcal{A}$ knots. A similar behavior is shown by the redshifted (NW) velocities, namely, The NW (positive distances) velocities shown smoothly rise and reach V$\sim 130$ km s$^{-1}$ at d$\sim 0.17$ kpc, then they rise upto V$\sim 259$ km s$^{-1}$ at d$\sim 1.18$ kpc, then decrese to a value V$\sim 216$ km s$^{-1}$ at d$\sim 1.8$ kpc, V$\sim 238$ kkm s$^{-1}$ at d$\sim 2.4$ kpc, and finally reach V$\sim 229$ km s$^{-1}$ at d$\sim 2.95$ kpc. Notice that the center of the slit the velocity is redshifted (by about V$\sim 34$ km s$^{-1}$).

\begin{figure}[bth]
\includegraphics[width=8cm,height=8cm]{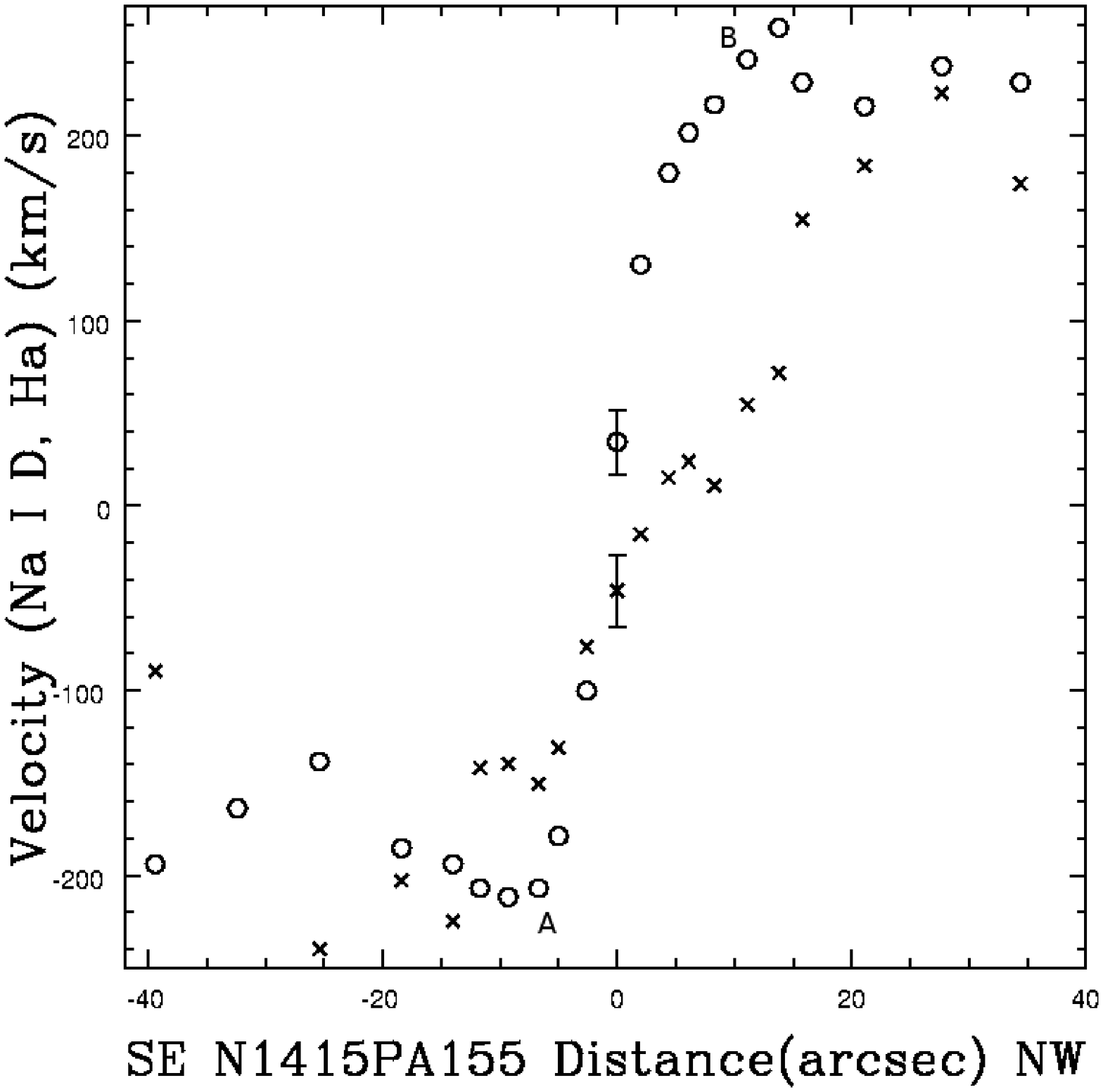}
\includegraphics[width=8cm,height=8cm]{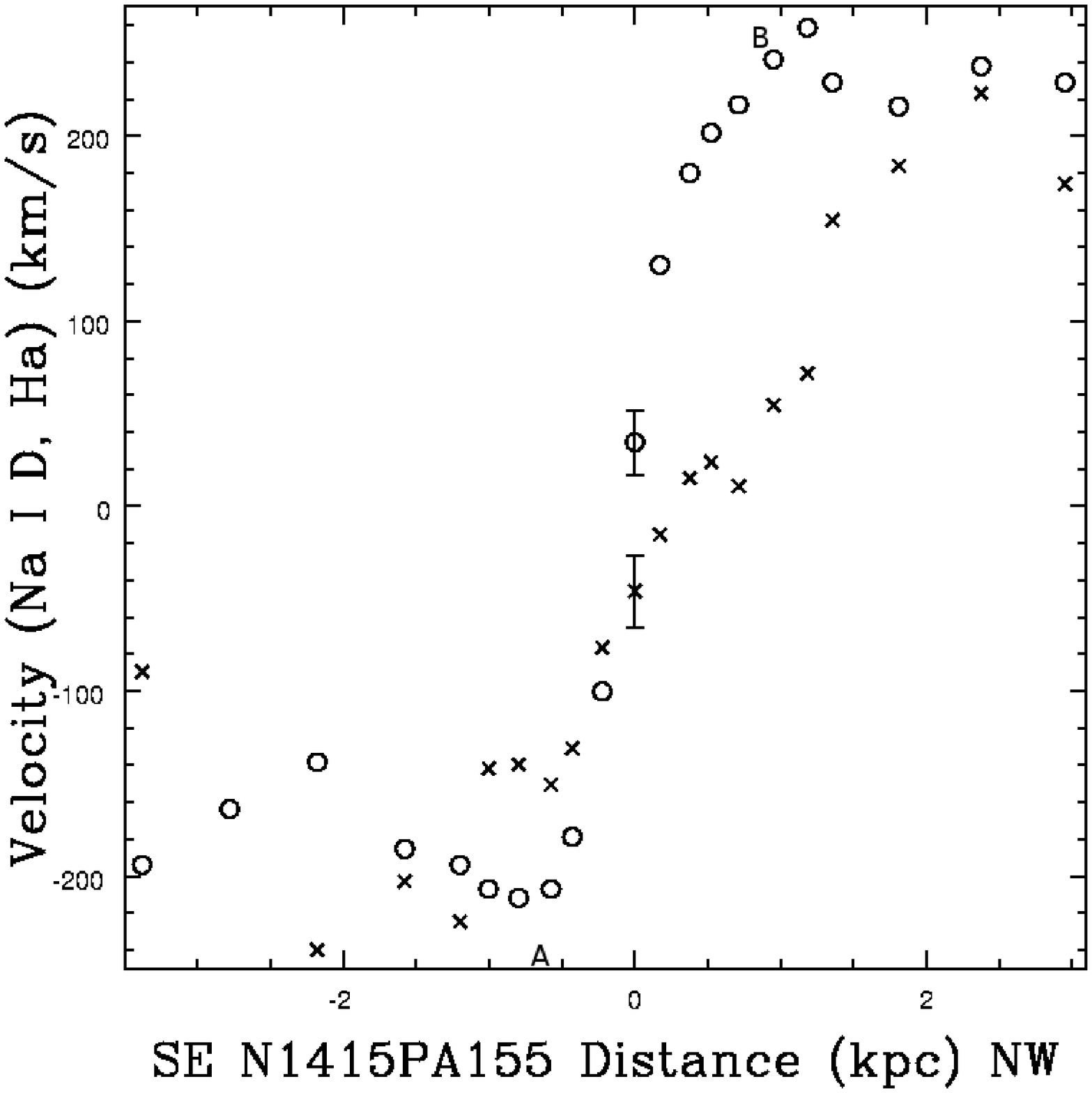}
\figcaption[figure=gdgnew22.ps]{Velocities versus distances (in arcsec in left plot) for the Na I D absorption lines (crosses) and simultaneously H$\alpha$ emission lines (open circles) for slit at P.A.$\sim 155\degr$. Velocities at d$ = 0\farcs0$ show the uncertainities for Na I D line $(+19,-19.5$ km s$^{-1}$, and for H$\alpha$ line $+17.2, -17.5$ km s$^{-1}$. Right plot shows the same as in left plot except that distance is in kpc. Notice the large difference in H$\alpha$ NW reshifted velocities (right in plots) compared to the Na I D velocities in the inner d$\sim 1.2$ kpc (d$\sim 14\farcs4$) and the difference shown in the SE blueshifted velocities (left on plots) specially at d$\sim 0.8$ kpc (d$\sim 9\farcs3$). Letter B on north axis, and A on bottom axis denote approximately the space locations of $\mathcal{B}$ and $\mathcal{A}$ knots. Notice that at the center of the slit the Na I D velocity is blueshifted by about V$\sim 46$ km s$^{-1}$, while the H$\alpha$ velocity is redshifted by about V$\sim 34$ km s$^{-1}$. \label{fig.5} }
\end{figure}

Fig. 5 (left plot shows velocity versus distance in arcsec, while right plot shows velocities versus distance in kpc) shows the Na I D velocities (crosses) from long slit at P.A.155$\degr$ (which is only 7$\degr$ from the photometric P.A.148$\degr$ of the disk of NGC 1415), simultaneously with H$\alpha$ heliocentric velocities (open circles).  The velocity values at R$ = 0$ show the uncertainities in both H$\alpha$ and Na I D lines. Letter B on north axis, and A on bottom axis denote approximately the space locations of $\mathcal{B}$ and $\mathcal{A}$ knots. Notice the higher H$\alpha$ velocities (in emission)  compared with the Na I D velocities (in absortion) at the locations shown by letter B and A. There is a large difference in H$\alpha$ (open circles) compared with the Na I D (crosses) in both the SE (left on plots) blueshifted velocities and the NW (right on plots) redshifted velocities in the sense that the absolute values show higher velocities for H$\alpha$ than Na I D. This difference is large in the redshifted velocities in the inner $0 \leq d \leq 1.2$ kpc (right side of plots), and in the blueshifted velocities in the inner $0.43 \leq d \leq 1.0$ kpc (left side of plots). $\mathcal{A}$ knot is located at a SE distance $\sim 755$ pc from the center of the galaxy, while $\mathcal{B}$ knot is located at a NW distance $\sim 884$ pc from the center of the galaxy.

\subsection{Rotation Curve in NGC 1415}

As briefly described in previous sections, our long slit spectroscopic observation detected lines of N[II] $\lambda 6548$ \AA, H$\alpha \lambda 6562.8$ \AA, [NII] $\lambda 6584$ \AA, [SII] $\lambda 6716$ \AA~ and [SII] $\lambda 6731$ \AA~in emission, while the convolved Na I D lines (spectrally unresolved due to our instrumental spectral sampling) $\lambda 5895,92$ \AA, and $\lambda 5889.95$ \AA~ were detected in absorption. Since the absorption Na I D line is optically thin, cold, residing most likely in the plane of the NGC 1415 galaxy, it most certainly shares the orbital motion of gas in the disk. We assume that the Na I D velocities versus distance from the slit at P.A.$\sim 155\degr$ shown in Fig. 3 and in Fig. 5 is representative of the innermost rotation curve in the barred galaxy NGC 1415.

\section{Geyser (Bipolar Outflow) from the Nucleus of NGC 1415?}

\subsection{Kinematics of $\mathcal{A}$ and  $\mathcal{B}$}

As mentioned earlier, the Na I D line velocities at the center of the spectrum at P.A.$\sim 155$ EofN in Fig. 3 and Fig 5  show negative velocity (blueshifted with respect to systemic) at distance 0.
	
Carefull observation of Fig. 3, Fig.4 and specially in Fig. 5 indicates that the H$\alpha$ velocity at SE distance from $0.4 \leq d \leq 1.5$ kpc are higher (blueshifted) than the Na I D line. The value of the Na I D line at about $\mathcal{A}$, d$\sim 800$ pc  (left side of plots in Fig.5), the velocity is V$(Na I)\sim 140$ km s$^{-1}$, while the value of the H$\alpha$ line is V$(H\alpha)\sim 212$ km s$^{-1}$. However, as we mentioned in the previous paragraph, the Na I D line (from Fig. 3, and Fig. 5) shows V$(Na I D line)_{d=0} \sim -46$ km s$^{-1}$ (blueshifted). If we were to force V$(Na I D line)_{d=0} \sim 0$ km s$^{-1}$, then we would be adding 46 km s$^{-1}$, that is, the new value of Na I D line at $\mathcal{A}$, d$\sim 800$ pc  would be V$(Na I D line) \sim 94$ km s$^{-1}$, and the difference is $\Delta V(H\alpha - Na I D line)_{SE} \sim 118$ km s$^{-1}$.
	
Similarly, the value of the Na I D line velocity at about $\mathcal{B}$, d$\sim 800$ pc (crosses at right side in Fig. 5), is V$(Na I D line)\sim 11$ km s$^{-1}$, while the value of the H$\alpha$ line is V$(H\alpha)\sim 217$ km s$^{-1}$. If we were to force V$(Na I D line)_{d=0} \sim 0$ km s$^{-1}$, then we would be adding 46 km s$^{-1}$, that is, the new value of Na I D line at $\mathcal{B}$, d$\sim 800$ pc would be V$(Na I D line) \sim 57$ km s$^{-1}$. Then the difference is $\Delta V(H\alpha - Na I D line)_{NW} \sim 160$ km s$^{-1}$. A value similar to the one obtained from $\mathcal{A}$.

We believe that the larger velocities seen in H$\alpha$ (compared with the rotation curve shown by the Na I D velocities in the disk of NGC 1415) suggest non circular kinematics in the sense that gas in $\mathcal{A}$ is moving away from the center (approaching us) and gas in $\mathcal{B}$ is moving away from the center (receding from us).

\subsection{Observational evidence of a H$\alpha$ Geyser (Bipolar Outflow) from the nucleus in NGC 1415}

Our estimated spatial location and kinematical analysis of $\mathcal{A}$ and $\mathcal{B}$ (left plot of Fig. 3, Fig. 4 and Fig. 5) indicates that H$\alpha$ gas shows larger (absolute) velocities compared with Na I D velocities. Thus $\mathcal{A}$ shows bluer velocities and thus gas is moving away from the nucleus (approaching us) at a velocity of about 118 km s$^{-1}$, while $\mathcal{B}$ shows redder velocities and thus gas is moving away from the nucleus (receding from us) a velocity of about 160 km s$^{-1}$. The P.A. of a line joining $\mathcal{A}$ with the nucleus and the P.A. of a line joining $\mathcal{B}$ with the nucleus are the same. Additionally, radio continuum radiation (20 cm, mostly synchrotron emission) has been detected with the VLA with FWHM$_{beam}8\farcs0$ angular resolution from the central region of NGC 1415 \citep{con90}. The central radio continuum map covers approximately the same area as our innermost H$\alpha$ emission and is elongated SE - NW at P.A.$\sim 162\degr$ very similar to the P.A. joining $\mathcal{A}$ and $\mathcal{B}$. Surprisingly the peak of the radio continuum emission does not come from the nucleus of NGC 1415, but instead it coincides with $\mathcal{A}$. This detection of apparently symmetric synchrotron 20 cm radio continuum emission at a similar P.A. as the line joining $\mathcal{A}$ and $\mathcal{B}$, together with our long slit kinematical data, suggests that this is a bipolar outflow from the nucleus, with the brighter knot $\mathcal{A}$ being in front of the galaxy and the fainter knot $\mathcal{B}$ being behind the disk. Polarization studies of radio continuum emission would be very useful to confirm this fact.

The kinematical age might be estimated by $\tau \sim d/v$, which for $d\sim 800$ pc and $v \sim 140$ km s$^{-1}$, $\tau \sim 5.6 $ Myrs, a very recent event in the life of a galaxy.

\subsection{Orientation of the P.A. of the outflow in comparison with P.A. of rotation axis of NGC 1415}

The P.A.$\sim 161\degr \pm 4\degr$ of the line joining  $\mathcal{A}$ and $\mathcal{B}$ knots is considerably different from the expected nuclear outflow perpendicular to the plane of a galaxy. If it were along the rotation axis of NGC 1415 (perpendicular to the plane), the expected outflow direction would be at P.A.$_{rotation-axis}\sim +238\degr$. Thus, the source of the bipolar outflow must be inclined to the plane of the disk of NGC 1415.

The unified model suggests that there is a black hole with an accretion disk and dusty torus associated with different types of AGN (Liners, being the low luminosity end of AGNs \citep{ant93}). The spatial location of $\mathcal{A}$ and $\mathcal{B}$ knots indicates that the accretion disk (which might be responsible of collimated bipolar outflow) is highly inclined $\sim 73\degr$ with respect to the plane of the disk galaxy. It is known that radio continuum emission P.A. off the nucleus does not necessarily agrees with photometric minor axis P.A. of host Seyfert galaxies \citep{sch97,kin00}. Different P.A. of the radio continuum emission from blobs straddling the nucleus have been detected with values different than the P.A. of the photometric minor axis of Sy 2 galaxies \citep{sch97,kin00}. One plausible interpretation of the orientation difference between nuclear bipolar outflow and rotation axis of the galaxy (at least in Seyfert galaxies) might be that it is a result of a minor galaxy merger \citep{nag99}.	

\subsection{Origin of the Nuclear Geyser (Bipolar Outflow) in NGC 1415}

Neither our angular resolution of the H$\alpha$ continuum-free image nor the angular resolutiion of the radio continuum emission is sufficient to resolve any long and narrow structure that links $\mathcal{A}$ or $\mathcal{B}$ knots with the center that could be identified as a jet \citep{con90}. So following the notation for the outflow in the center of the disk galaxy M101 \citep{moo95}, we may call the cause of the outflow a geyser.

In the barred galaxy NGC 1415, could the geyser be caused by a recent activity of a low mass extragalactic black hole? or by a compact burst of thousands of O, B star formation ? Nuclear compact bursts of thousands of O, B star formation give rise to bubbles that are usually perpendicular to the plane, that is, parallel to the rotation axis of the disk \citep{car90,vei94}\footnote{In the case that the outflow from NGC 3079 parallel to the rotation axis were due to an AGN, the black hole mass would be M(NGC 3079$_{BH} \sim 8.4 \times 10^7$ M$_{\odot}$, see section below.}, and thus is not a plausible source in NGC 1415 because the outflow P.A. is quite different than the P.A. of the rotation axis of the disk.

It does seem that the barred galaxy NGC 1415 is one of several {\it normal} disk galaxies with weak  optical and radio continumm emission from its nucleus. Other normal galaxies are M51 (NGC 5194 Sbc(s)I-II) \citep{for85,cra92,cec88}, M81 (NGC 3031, SA(s)ab) \cite{goa76}, M101 (NGC 5457 SAB(rs)cd I) \citep{moo95}, NGC 3367 (SBc(s)II) \citep{gar98,gar02}, see below.

Studies of the innermost central region in M51 of radio continuum VLA high resolution mapping \citep{for85,cra92} indicate the presence of a radio continuum nuclear source (with spectral index $\alpha \sim -0.67$ indicating non-thermal emission), a jet and two radio continuum bubbles straddling the nucleus \citep{for85}. M51 has an inner bar with a$_{bar} \sim 20''$ at a P.A.$\sim 135\degr$ \citep{pie86} (832 pc, at a distance of 8.58 Mpc \citep{mcq16}). Additionally, reported optical spectroscopy and Fabry Perot H$\alpha$+[N II 6584] \citep{cec88} show the spatial distribution and kinematics of the innermost central ionized gas in M51 \citep{cec88}. The optical emission spatial distribution is very similar to the radio continuum emission, where the southern bubble seems to be a working surface (of a nuclear outflow) moving at $\sim 200 -- 500$ km s$^{-1}$ \citep{cec88}.

16 Spectograms were centered near H$\alpha$ and [N II 6583.4] and the slits always intersected the nucleus of M81 \citep{goa76}. The nucleus contains a point source radio continumm source (non-thermal) from a diameter less than $2\farcs0$ ($\sim 31$ pc) \citep{deb76}. The kinematics of the ionized gas in the innermost central region of M81 reveals gas (with non-circular velocities) with {\it nuclear outflow} velociy of about $v_{outflow} \sim 38$ km s$^{-1}$.

Another example of two H$\alpha$, continuum-free, knots have been reported straddling the nucleus, in the north-south direction, from the disk galaxy M101 \citep{moo95}. Several long spectroscopic slits observations of the innermost central region of M101 reveal that kinematics of the two H$\alpha$ knots straddling the nucleus is most likely a geyser or bipolar outflow with velocity less than 100 km s$^{-1}$ \citep{moo95}.

The barred galaxy NGC 3367 (SBc) shows bipolar radio continuum 20cm and 6cm emissions from interferometric observations with synchrotron emission lobes extending upto 6 kpc straddling the compact nucleus \citep{gar98,gar02}.

All of these examples of weak optical and radio continuum emissions from the nuclei of {\it normal} disk galaxies may be explained by having an active nucleus with a low mass black hole.

With the relation M$_{BH}$ versus $\sigma_*$, by now well accepted for extragalactic massive black holes \citep{geb00,mer01a,mer01b} one may estimate the masses of the each black hole of the {\it normal} disk galaxies briefly described earlier. The values of their  central velocity dispersions $\sigma_*$ are taken from Ho et al. (2009). The mass is obtained from the expresion M$_{BH} \sim 1.2 \times 10^8$ M$_{\odot}$ $(\sigma_*/200 km/s)^{3.75}$ \citep{geb00}.

Thus, M(M51)$_{BH} \sim 7.6 \times 10^6$ M$_{\odot}$,  M(M81)$_{BH} \sim 5.4 \times 10^7$ M$_{\odot}$, M(M101)$_{BH} \sim 3.97 \times 10^4$ M$_{\odot}$, and M(NGC3367)$_{BH} \sim 1.4 \times 10^6$ M$_{\odot}$.

As a comparison, NGC 5548 a Seyfert 1.5 disk galaxy (Hubble type R$'$SA(0)/as) shows a unresolved radio continuum source from the nucleus (at $0\farcs25$ angular resolution) and two unresolved radio continuum sources  straddling the nucleus at d$\leq 2\farcs0$ (d $\leq 1.4$ kpc \citep{kuk95}), shows X-ray variable emission and it has been detected as a X-ray warm absorber. M(NGC5548)$_{BH} \sim 5 \times 10^8$ M$_{\odot}$ is a more massive black hole. It is the result of a galaxy merger event.

\begin{figure}[h]
\includegraphics[width=10cm,height=10cm]{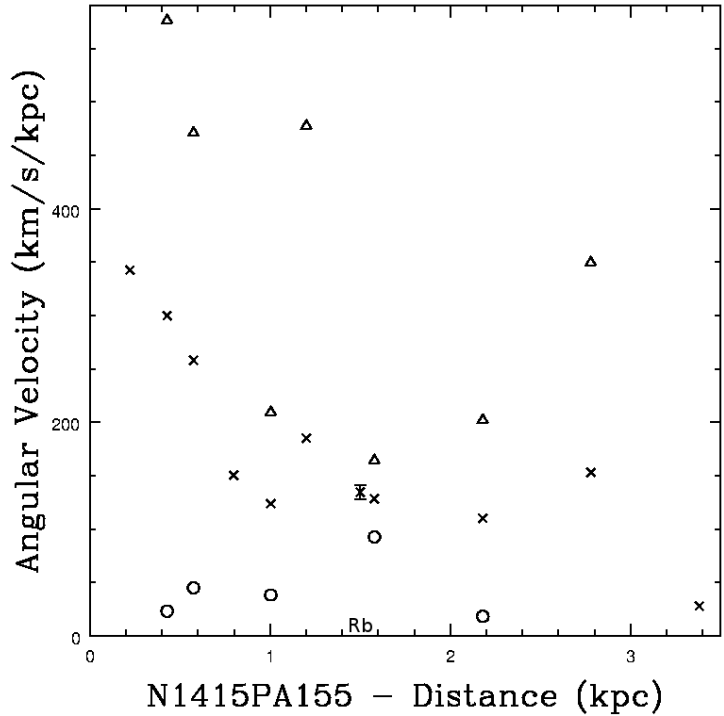}
\figcaption[figure=gdgnew21.eps]{Plot of angular velocities from the blueshifted velocity's Na I D rotation curve (see Fig. 3 left plot) $\Omega_{gas}$ (crosses), $\Omega_{gas} - \kappa/2$ (open circles), and $\Omega_{gas} + \kappa/2$ (open triangles). With $\mathcal{R} = 1$, the radius where CR occurs at the end of the stellar bar, a$_{bar} \sim 1.5$ kpc \citep{gar00} (letters Rb are shown on the bottom axis), $\Omega_{gas} \sim 134 \pm  7$ km s$^{-1}$ kpc$^{-1}$, and thus $\Omega_{bar} \sim 134 \pm7$ km s$^{-1}$ kpc$^{-1}$. This value is shown at R$\sim 1.5$ kpc with a cross with error bars. This value is higher than other values from ellipsoidal stellar bars. It is higher than $\Omega_{gas} - \kappa(r)/2$, and does not cross $\Omega_{gas} + \kappa(r)/2$.
\label{fig. 6}}
\end{figure}

\section{Angular Velocity of the stellar bar, $\Omega_{bar}$, in NGC 1415}

The angular velocity of a stellar bar ($\Omega_{bar}$) in a SB galaxy is among the most important parameters that govern the galaxy's internal dynamics, kinematics and the morphology of its internal structure \citep{bin87}. Hubble early type barred galaxies tend to have flat light profiles, while Hubble late type SB galaxies tend to have exponential light profiles along the ellipsoidal bars \citep{com93,elm96}. For slightly non circular star and gas orbits perturbed by a non-axisymmetrical gravitational potential, orbits have natural resonant frequencies. If the gravitational field generated by the stellar bar perturbs a gas orbit at or near one of its resonant frequencies, then the response of the orbit will be large.

\subsection{Methods for estimating $\Omega_{bar}$ from gas kinematics}

One method to estimate the angular pattern speed ($\Omega_{bar}$) for stellar bars is through the resonances method \citep{bin87,elm96}, where it is based on gas kinematics, the locations of resonance rings, and ILR and OLR structures \citep{bin87,elm96}. Another method for estimating a bar pattern speed is that proposed by Tremaine \& Weinberg (1984).

In this paper we will utilize the first method. Resonances occur when the circular angular velocity $\Omega_{gas}$ and the radial epicycle frequency, $\kappa(R)$, in the unperturbed orbit satisfies one of the three conditions: $\Omega_{bar} = \Omega_{gas}$ at CR, ($\mathcal{R} = R_{CR} / R_{bar}$), $\Omega_{bar} = \Omega_{gas} - \kappa /2$ (near an ILR), and $\Omega_{bar} = \Omega_{gas} + \kappa /2$ (near an OLR) \citep{lyn72,con80a,con80b,bin87}.

Gas outside CR drifts towards near an OLR, while gas inside CR drifts towards near an ILR \citep{lyn72,con80a,con80b,bin87}. The radial epicycle frequency in terms of the effective gravitational potential is given by the expression $\kappa^2(R) = \left(\partial^2 (\Phi_{eff})/ \partial (R^2) \right)$, where $\Phi_{eff}$ is the sum of the gravitational potential energy of the orbiting gas and the kinetic energy associated with its motion in the azimuthal direction \citep{bin87}. In terms of $\Omega_{gas}$, $d(\Omega_{gas})/dR$, the radial epicycle frequency can be estimated by the expression $\kappa^2(R) = \left( 4\Omega^2(R)_{gas} + 2R\Omega(R) (d\Omega(R)_{gas})/dR \right)$ \citep{bin87}. The derived observed values of $\Omega(Na I D lines)_{gas}$ (crosses), $\Omega_{gas} - \kappa /2$ (open circles), and $\Omega_{gas} + \kappa /2$ (open triangles) are shown in Fig. 6.

In a disk galaxy with an additional non-axisymmetric gravitational potential as is a stellar bar, CR, $\Omega_{bar} = \Omega_{gas}$, occurs at the end of the ellipsoidal stellar bar ($\mathcal{R} = 1$) \citep{con80a,con80b}. In the disk barred galaxy NGC 1415, R$_{boxy bar} \sim 1.5$ kpc \citep{gar00}, then from Fig. 6, $\Omega_{gas}$ at R$ \sim 1.5$ kpc has a value of $\Omega_{gas} \sim 134 \pm 7$ km s$^{-1}$ kpc$^{-1}$ and thus $\Omega_{bar} \sim 134$ km s$^{-1}$ kpc$^{-1}$. $\Omega_{bar}$ does not cross the derived $\Omega_{gas} - \kappa(R)/2$, nor $\Omega_{gas} + \kappa(R)/2$. We notice, from optical continuum image (see contours in Fig.1) that at a distance just at the end of the boxy bar ($\sim 2$ kpc) there are elongated isophotes indicating the start of the inner spiral arms (see Fig 1). It is difficult to tell at which distance OLR were to occur, since the values of $\Omega_{gas}$ changed rapidly at large distances and $\kappa(R)$ was thus not easy to estimate. This constant value of $\Omega_{bar}$ for this boxy bar in NGC 1415 is very large at least a factor from 3 to 5 with respect to values of other ellipsoidal bars \citep{her05,fat07,fat09}. From numerical simulations, large values of $\Omega_{bar}$ are expected for an {\it inner} bar (see computer simulated model 2 values of doubly barred galaxies \citep{mac00}). Is the large NIR bar in NGC 1415 \citep{lau11} the primary bar?

This value of $\Omega_{bar} \sim 134 \pm 7$ km s$^{-1}$ kpc$^{-1}$ in NGC 1415 is large, however it is similar to the estimated large values in NGC 4303 for $\Omega_{bar}$ when $\mathcal{R} = 1$ \citep{gar19}\footnote{In the case of NGC 4303, if one were to take $a_{bar} = 20\farcs$ ($\sim 1.5$ kpc) \citep{mar95} and $\mathcal{R} =1$, then $\Omega_{bar} \sim 160$ km s$^{-1}$ kpc$^{-1}$ (CR $\sim 1.5$ kpc), it would not cross $\Omega_{gas} - \kappa(R)/2$, but it would cross $\Omega_{gas} + \kappa(R)/2$ at about 2.4 kpc (OLR) which coincides with the location of the inner southern spiral arm in NGC 4303 \citep{gar19}. While if one were to take $a_{bar} \sim 28\farcs6$ ($\sim 2.1$ kpc) \citep{gad06} and $\mathcal{R} =1$, then $\Omega_{bar} \sim 120$ km s$^{-1}$ kpc$^{-1}$ (CR $\sim 2.1$ kpc), it also does not cross $\Omega_{gas} - \kappa/2$, but it crosses $\Omega_{gas} + \kappa /2$ at R$_{OLR} \sim 2.7$ kpc which would approximately also coincide with the spatial location of the inner southern spiral arm in NGC 4303 \citep{gar19}.}.

Notice that the bright inner spiral arms [at about d$\sim 40''$, or d$\sim 3.43$ kpc] are in the inner bright part of the inner disk of NGC 1415, and the disk is much larger, as reported in NED, with ESO-LV quick blue IIa-O semi major axis extending to 169$\farcs4$ or 14.53 kpc.

\subsection{Comparison of $\mathcal{R}$ in NGC 1415 with values from other galaxies}

How does the ratio $\mathcal{R} = 1$ in NGC 1415 compare with the values in other barred galaxies? Seven out of nine Hubble early type SB galaxies,  $\mathcal{R} = 1.2 \rightarrow 2.2$ \citep{elm96}. In a model-based study of 38 SB galaxies (using NIR and optical images from the OSUBGS), $\mathcal{R}$ is near 1.15 in types SB0/a-SBab \citep{rau08}. Values have been obtained for $\Omega_{bar}$ from 10 barred galaxies with large ellipsoidal stellar bars and $0.8 \leq \mathcal{R} \leq 1.1 $. In a study of 15 CALIFA SB galaxies, using the TW method, $\langle \mathcal{R} = 1.3 \rangle$ for SB0-SB0/a galaxies \citep{agu15}. 

Stellar bars observed almost edge on may show inner boxy optical isophotes (for example NGC 1415 \citep{gar96,gar00}, and NGC 4569 \citep{2ma03,gar19}). Yet there is no statistical observation, to our knowledge, of values of $\Omega_{bar}$ for these boxy small inner bars. In NGC 1415 the stellar bar is boxy with R$_{bar} \sim 1.5$ kpc small as compared with the radius of the large disk blue IIa-O semi major axis of 14.5 kpc, $a_{bar}/R_{25} \sim 0.1$ which is smaller than any value reported from other barred galaxies from the S$^4G$ 3.6$\mu m$ \citep{dia16}.

\section{Summary and Conclusions}

Our previous imaging H$\alpha$ continuum-free data from the central $\pm 20\farcs0$ of the barred galaxy NGC 1415 showed a) two central bright H$\alpha$, $\mathcal{A}$, and $\mathcal{B}$ knots neither of them coinciding with the optical continuum nucleus, b) emission from regions around the nucleus, and c) emission from knots in the SE and NW inner spiral arms. From our previous optical red continumm (filter {\it I} $\lambda 8040$ \AA$~$ a mass distribution model was made which included a two stellar bars: one to reproduce the observed inner ellipsoidal isophotes (R$_{bar I} \sim 1$ kpc) and a second to reproduce the boxy-shaped isophotes (R$_{bar II} \sim 1.5$ kpc) \citep{gar00}.

In this study we have obtained kinematical data from the disk of NGC 1415 with a long slit spectrum at  P.A.$\sim 155\degr$. We were able to detect the convolved Na I D lines in absorption, [NII] $\lambda 6548, 6584$, H$\alpha \lambda 6562.8$, and [SII] $\lambda 6716, 6731$ lines in emission.
	
We estimated that the heliocentric velocities of Na I D absorption lines from a long slit spectrum at P.A.+155$\degr$ (which is closest to the photometric P.A. +148$\degr$ of the disk of NGC 1415) versus distance [taken as V$_{sys} \sim 1564$ km s$^{-1}$]  may be representative of the rotation curve in NGC 1415.
	
From a comparison of the heliocentric velocities of Na I D absorption lines and H$\alpha$ emission lines we observed an excess in velocity of the H$\alpha$ emission lines associated with $\mathcal{A}$ (bluer velocities) and $\mathcal{B}$ (redder velocities) knots as compared with the velocities shown by the Na I D lines at the same distances, $|\Delta V| \sim 140$ km s$^{-1}$.

From reported VLA radio continuum (20 cm) mapping of the inner $30''$, with an angular resolution of $8''$ \citep{con90}, the peak of the 20 cm radio continuum emission surprisingly does not coincide with the nucleus but instead it coincides with $\mathcal{A}$.

Our interpretation of the spatial location (very symmetric the positions of $\mathcal{A}$ and $\mathcal{B}$) and blueshifted velocities of $\mathcal{A}$, and redshifted velocities of $\mathcal{B}$ and the detected radio continuum 20 cm emission at an angular resolution of $8\farcs0$(synchrotron), is that they are the result of gas moving away from the nucleus. We interpret that there may be a geyser (bipolar outflow from the nucleus) with V$_{geyser} \sim 140$ km s$^{-1}$. The geyser lies at a P.A.+165$\degr$ which is approximately 73$\degr$ from the rotation axis of the disk of NGC 1415, namely it is not perpendicular to the disk of NGC 1415. The cause of the geyser could be a low mass black hole with an accretion disk highly inclined to the plane of the disk of NGC 1415 (similar nuclear activity from other normal disk galaxies has been reported in M51, M81, M101, and NGC 3367 that may be explained by the existence of low mass black hole in each of them).

From our estimated rotation curve we were able to estimate the angular velocity of the gas, $\Omega_{gas}$, the radial epicycle frequencies $\kappa(r)$, $\Omega_{gas} - \kappa/2$, and $\Omega_{gas} + \kappa/2$.

A value of $\Omega_{bar}$ is considered in the barred galaxy NGC 1415 with $\mathcal{R} = 1$, that is, CR is at 1.5 kpc (radius of the fitted boxy stellar bar), with an estimated value $\Omega_{bar} \sim 134$ km s$^{-1}$ kpc$^{-1}$. This value of $\Omega_{bar}$ is high and does not cross $\Omega_{gas} - \kappa /2$, nor $\Omega_{gas} + \kappa /2$.

\acknowledgments{ We would like to thank constructive comments to improve this version of the text to an anonymous referee. We also would like to thank for their useful comments, suggestions and references to previous published articles related to lengths and angular velocities of stellar bars to F. Combes (France), E. Athanassuola (France), and B. Elmegreen (USA). This work has made extensive use of the NASA/IPAC Extragalactic Database (NED) which is operated by the Jet Propulsion Laboratory, California Institute of Technology, under contract with the National Aeronautics and Space Administration. We also would like to thank the operator control of the 2.1m GH telescope in Cananea, M\'exico for his help with the observations in December of 2000. This research has made use of the VizieR catalogue access tool, CDS, Strasbourg, France (DOI:10.26093/cds/vizier). The original description of the VizieR service was published in A\&AS, 143, 23, and the astronomy package IRAF (developed by NOAO. NOAO is managed by the Association of Universities for Research in Astronomy under Cooperative Agreement with the National Science Foundation, USA).}

\appendix
\section{Neutral Sodium Optical interstellar Absoprtion/Emission Lines}

Sodium atom (Na) has 11 protons, 11 neutrons and 11 electrons. Na I first ionization potential is 5.14 eV. The electron distribution in the sodium atom is Na I [{$1S^2 2S^2 2P^6$} 3$S^1$]. The neutral sodium {\it D} lines in absorption arise from the hyperfine structure from the levels $3 ^2S \rightarrow 3 ^2P$ \citep{mei37}. In 1937 the transition $3 ^2S_{1/2} \rightarrow 3 ^2P_{1/2}$ reported a wavelength $\lambda_{D1} \sim 5895.932$ \AA and was denoted as D$_1$ line, while the transition $3 ^2S_{1/2} \rightarrow 3 ^2P_{3/2}$ has the wavelength $\lambda_{D2} \sim 5889.965$ \AA and was denoted as D$_2$ line. More recent work show that the wavelength $\lambda_{D1} \sim 5895.924$ \AA, while the transition $3 ^2S_{1/2} \rightarrow 3 ^2P_{3/2}$ has the wavelength $\lambda_{D2} \sim 5889.951$ \AA \citep{mcn63}. The energy difference between the upper and lower states in the Na D$_1$ line is $\Delta E_{D1} \sim 2.103$ eV, while it is $\Delta E_{D2} \sim 2.105$ eV. They are the so called unresolved D$_1$, and D$_2$ lines of neutral sodium, Na I D \citep{mcn63}. The wavelength separation between the Na I D$_1$ line and the Na I D$_2$ line is $\Delta \lambda \sim 5.973$ \AA.

Galactic and extragalactic Na I D lines are detected in absorption, where the underlying continuum is strong, in the interstellar medium \citep{spi78,gra79,dys80}. The cosmic composition of Na, in our galaxy, is $12 + [log N_{Na}/N_H] \sim 6.25 \rightarrow 6.3$ \citep{spi78,dys80}. Studies in our galaxy indicate that Na I abundance is proportional to the square of the local hydrogen density, and thus it will be concentrated in the colder, denser portions of the clouds \citep{wel94}. Neutral hydrogen, HI, and sodium, Na I, are constituents of diffuse neutral clouds at a typical temperature of about $70\degr \rightarrow 80\degr$ K. These clouds also contains ions such as C$^+$, Ca$^+$, etc. produced by photoionization by starlight \citep{spi78,dys80}. The mean electron density is about 0.1 cm$^{-3}$, corresponding to $n_H \sim 10^3$ cm$^{-3}$ (Ca atom, in general, is depleted by a factor of 100 to 1000 relative to Na atom \citep{spi78}). The median Na I column density, from observation of absorption towards 38 bright stars in our galaxy is $log N[(Na I)] \sim 11.09$ \citep{wel94}, while it is $log N[(Na I)] \sim 12.8$ towards $\alpha$ Cygni line of sight \citep{bla80}. If neutral sodium, Na I, and neutral hydrogen, HI, are characterized by the same temperature and turbulent velocity in diffuse clouds, the optical depth $\tau$ is less than 1, namely, the Na I D absorption lines are optically thin \citep{wel94}. Na I D lines have been detected in absorption towards the galaxy NGC 5128 (Cen A) \citep{gra79}, where it is reported  that within the errors of measurements, the wavelenghts are redshifted, in Na I D lines, the same as those measured for NGC 5128 emission lines at each point, and it is clear that the absorbing material shares the motion of the gas which is responsible for the emission lines.

\vspace{5mm}
\facility{Haro Observatory, OAGH(2.1m) Cananea, M\'exico}

\end{document}